\documentstyle{aipproc}
\begin{document}     
\begin{center}{\Large\bf Lengths of the First Days of the Universe}\end{center}
\begin{center}{Moshe Carmeli}\end{center}
\begin{center}{Department of Physics, Ben Gurion University, Beer Sheva 84105, 
Israel}\end{center}
\begin{center}{Email: carmelim@bgumail.bgu.ac.il}\end{center}
\begin{abstract}
The early stage of the Universe is discussed and the time lengths of its first 
days are given. If we denote the Hubble time in the zero-gravity limit by 
$\tau$ (approximately 12.16 billion years), and $T_n$ denotes the length of 
the $n$-th day, then we have the very simple relation $T_n=\tau/(2n-1)$. Hence 
we obtain for the first days the following lengths of time: $T_1=\tau$, 
$T_2=\tau/3$, $T_3=\tau/5$, etc.
\end{abstract}
In this Note we calculate the lengths of days of the early
Universe, day by day, from the first day after the Big Bang on up to our 
present time. We find 
that the first day actually lasted the Hubble time in the limit of zero 
gravity. If we denote the Hubble time in the zero-gravity limit by $\tau$ 
which equals about 12.16 billion years and $T_n$ denotes the length
of the $n$-th day in units of times of the early Universe, then we have a very
simple relation 
$$T_n=\frac{\tau}{2n-1}.\eqno(1)$$
Hence we obtain for the first few days the following lengths of time:
$$T_1=\tau,\hspace{5mm}T_2=\frac{\tau}{3},\hspace{5mm}T_3=\frac{\tau}{5},
\hspace{5mm}T_4=\frac{\tau}{7},\hspace{5mm}T_5=\frac{\tau}{9},\hspace{5mm}
T_6=\frac{\tau}{11}.\eqno(2)$$

It also follows that the accumulation of time from the first day to the second,
third, fourth, ..., up to now is just exactly the Hubble time. The Hubble time
in the limit of zero gravity is the maximum time allowed in nature. 

Using 
Cosmological Special Relativity [1-4], the calculation is very simple. We 
assume that the Big Bang time with respect to us now was $t_0=\tau$, the time 
of the first day after that was $t_1$, the time of 
the second day was $t_2$, and so on. In this way the
time scale is progressing in units of one day (24 hours) in our units of 
present time. The time difference between $t_0$ and $t_1$, denoted by $T_1$, 
is the time as measured at the early Universe and is by no means equal to one
day of our time. In this way we denote the times elapsed from the Big Bang to 
the end of the first day $t_1$ by $T_1$, between the first day $t_1$ and the 
second day 
$t_2$ by $T_2$ and so on. According to the rule of the addition of cosmic times one 
has, for example,
$$t_6+1(day)=\frac{t_6+T_6}{1+t_6T_6/\tau^2}.\eqno(3)$$
A straightforward calculation then shows that
$$T_6=\frac{\tau^2}{\tau^2-\left(\tau-6\right)\left(\tau-5\right)}=
\frac{\tau^2}{11\tau-30}.\eqno(4)$$

In general one finds that
$$T_n=\frac{\tau^2}{\tau^2-\left(\tau-n\right)\left(\tau-n+1\right)},\eqno(5)$$ 
or
$$T_n=\frac{\tau}{n+\left(n-1\right)-n\left(n-1\right)/\tau}.\eqno(6)$$
As is seen from the last formula one can neglect the last term in the 
denominator in the first approximation and we get the simple Eq. (1).

From the above one reaches the conclusion that the age of the Universe
exactly equals the Hubble time in vacuum $\tau$, i.e. 
12.16 billion years, and it is a universal constant [6]. This means that the 
age of the Universe tomorrow will be the same as it was yesterday or today. 

But this might not go along with our intuition since we usually deal with short
periods of times in our daily life, and the
unexperienced person will reject such a conclusion. Physics,
however, deals with measurements. 

In fact we have exactly a similar situation
with respect to the speed of light $c$. When measured in vacuum, it is 300
thousands kilometers per second. If the person doing the measurement tries to 
decrease or increase this number by moving with a very high speed in the 
direction or against the direction of the propagation of light, he will find
that this is impossible and he will measure the same number as before. The
measurement instruments adjust themselves in such a way that the final result
remains the same. In this sense the speed of light in vacuum $c$ and the 
Hubble time in vacuum $\tau$ behave the same way and are both universal 
constants.

The similarity of the behavior of velocities of objects and those of cosmic 
times can also be demonstrated as follows. Suppose a rocket moves with  the
speed $V_1$ with respect to an observer on the Earth.
We would like to increase that speed to $V_2$ as measured by the observer on 
the Earth. In order to achieve this, the
rocket has to increase its speed not by the difference $V_2-V_1$, but by 
$$\Delta V=\frac{V_2-V_1}{1-V_1V_2/c^2}.\eqno(7)$$
As can easily be seen $\Delta V$ is much larger than $V_2-V_1$ for velocities
$V_1$ and $V_2$ close to that of light $c$.
This result follows from the rule for the addition of 
velocities,
$$V_{1+2}=\frac{V_1+V_2}{1+V_1V_2/c^2},\eqno(8)$$ 
a consequence of Einstein's famous Special Relativity Theory [5].
In cosmology, we have the analogous formula
$$T_{1+2}=\frac{T_1+T_2}{1+T_1T_2/\tau^2}\eqno(9)$$
for the cosmic times.

\end{document}